\documentclass[a4paper,longbibliography,twocolumn,superscriptaddress,showkeys,pre, aps, floatfix, nofootinbib]{revtex4-1}

\usepackage[utf8]{inputenc}
\usepackage[english]{babel}
\usepackage{amssymb}
\usepackage{amsmath}
\usepackage{graphicx}
\usepackage[caption=false]{subfig}
\usepackage{float}
\usepackage[normalem]{ulem}
\usepackage{color}

\definecolor{myred}{rgb}{0.6,0,0}
\definecolor{myblue}{rgb}{0,0,0.6}
\definecolor{sigcol}{RGB}{17,97,165}
\definecolor{shadecolor}{RGB}{211,220,238}
\definecolor{edgecolor}{RGB}{17,97,165}

\newcommand{\km}{\left<k\right>}

\newcommand{\eqcomma}{\,,}

\newcommand{\gvec}[1]{\boldsymbol{#1}}

\renewcommand{\subsubsection}[1]{}

\begin{document}

\author{Kaj-Kolja Kleineberg}
\email{kajkoljakleineberg@gmail.com}
\affiliation{Departament de F\'isica Fonamental, Universitat de Barcelona, 
Mart\'i i Franqu\`es 1, 08028 Barcelona, Spain}
\author{Dirk Helbing}
\email{dirk.helbing@gess.ethz.ch}
\affiliation{
Computational Social Science, ETH Zurich, Clausiusstrasse 50, CH-8092 Zurich, Switzerland}
\date{\today}

\title{A ``Social Bitcoin'' could sustain a democratic digital world}

\begin{abstract}
A multidimensional financial system could provide benefits for individuals, companies, and states. Instead of top-down control, which is destined to eventually fail in a hyperconnected world, a bottom-up creation of value can unleash creative potential and drive innovations. 
Multiple currency dimensions can represent different externalities and thus enable the design of incentives and feedback mechanisms that foster the ability of complex dynamical systems to self-organize and lead to a more resilient society and sustainable economy. Modern information and communication technologies play a crucial role in this process, as Web 2.0 and online social networks promote cooperation and collaboration on unprecedented scales. 
Within this contribution, we discuss how one dimension of a multidimensional currency system could represent socio-digital capital (Social Bitcoins) that can be generated in a bottom-up way by individuals who perform search and navigation tasks in a future version of the digital world. 
The incentive to mine Social Bitcoins could sustain digital diversity, which mitigates the risk of totalitarian control by powerful monopolies of information and can create new business opportunities needed in times where a large fraction of current jobs
is estimated to disappear due to computerisation.
\end{abstract}


\maketitle


\section{Modern socio-economic challenges require a new approach}

Nowadays we are facing a number of serious problems such as financial instabilities, an unsustainable economy and related global warming, the lack of social cooperation and collaboration causing the rise of conflict, terrorism and war. 
Traditional approaches to remedy such problems are based on top-down control. 
Whereas in the past this way of thinking worked reasonably well, 
the high interconnectivity in modern systems will eventually but unavoidably lead to its failure
as systems become uncontrollable by central entities due to stronger internal effects, leading to often unpredictable cascading behavior~\cite{helbing2013globally} and catastrophic failures~\cite{BPPSH10}.

Instead of entirely top-down based approaches, designing mechanisms to promote desired results like increased cooperation, coordination, and better resource efficiency could help to deal with current socio-economic challenges. Importantly, a multidimensional incentive system is needed to design the desired interactions and appropriate feedback mechanisms~\cite{interaction:support,new:economy}. Such incentives have to be implemented in a bottom-up way, allowing systems to self-organize~\cite{helbing:self} and thus promoting creativity, innovation, and diversity~\cite{blog:self-organized:society}. 

Diversity acts as a motor of innovation, can promote collective intelligence~\cite{page:difference}, and is fundamental for the resilience of society~\cite{2016man,change:complex}. This renders socio-economic and cultural diversity equally important as biodiversity. 
The importance of diversity, however, is not restricted to individual, cultural, social, or economic domains. For instance, diversity among digital services in competition for the attention of users can mitigate the risk of totalitarian control and manipulation by extremely powerful monopolies of information. 
As we explain in Sec.~\ref{sec_model}, the loss of diversity in the digital world can lead to a systematic and irreversible collapse of the digital ecosystem~\cite{ecology20,worldmodel}, akin 
to the loss of biodiversity in the physical ecosystem. 
Such a collapse can have dramatic consequences for the freedom of information and eventually for the freedom of society.
In this contribution, we show how such a catastrophic collapse could be avoided on a systematic level by introducing a multidimensional incentive system in which an appropriately designed cryptocurrency provides an incentive for individuals to perform certain tasks in their socio-digital environment. We refer to this cryptocurrency as ``Social Bitcoins''\footnote{The details of the implementation of such a cryptocurrency is beyond the scope of this contribution.}.

Importantly, to successfully meet these challenges, tools, ideas and concepts from complexity science have to be combined with technologies like the blockchain, economic knowledge (and potentially Internet of Things technology to measure ``externalities'').

\section{A multidimensional financial system}

The invention of money has led to unprecedented wealth and has provided countless benefits for society. 
However, the current monetary system is not appropriate any more to control highly interconnected dynamical complex systems like the ones our economy and financial system nowadays form. Whereas such systems are in general difficult to control and understand and nearly impossible to predict, they exhibit the tendency to self-organize~\cite{helbing:self,PhysRevLett.59.381}. New approaches to face todays challenges should therefore take advantage of this system intrinsic tendency. 

Central banks like the ECB can control the amount of money in the market by means ranging from adjusting interest rates to quantitative easing. Recently, the ECB has lowered interest rates to the lowest value of all time (even introducing negative rates for some bank deposits~\cite{reuters:negative}) and has further increased its efforts to buy government bonds~\cite{quantitative:easing}. 
These measures are intended to boost economy and increase inflation in the Euro zone to the target of $2\%$. Despite these efforts, inflation has remained close to $0\%$, raising doubts about the capacity to act and the credibility of the ECB~\cite{reuters:credibility}.
Furthermore, liquidity pumped into the market does not reach efficiently enough the real economy.
As a consequence recently ``helicopter money'' has been discussed as a possible solution~\cite{reuters:helicopter,helicopter}. 
Importantly, these problems are not limited to the Euro zone.
For example, due to the interconnected nature of our economic and financial systems, the state of the global economy limits the decisions the Fed can take concerning a raise of interest rates, as such a raise could pose a threat for the global economy~\cite{reuter:fed}.

The problem is that the current monetary system provides only a one-dimensional control variable. 
Let us consider the human body and how it self-organizes as an example. 
To ensure its healthy function, it is not enough to adjust only the amount of water one drinks. Instead, the body needs water, air, carbohydrate, different proteins and vitamins, mineral nutrients and more. None of these needs can be replaced by another. 
Why should this be different in systems like our economy, the financial system, or society?

Indeed, a multidimensional currency system could help to solve the problems mentioned above, where the different dimensions can be converted at a low (or negligible) cost.
Such multidimensional incentive system could be used to promote self-organization of financial and economic systems in a bottom-up way~\cite{thinking:ahead}. 
This opens the door to ``Capitalism 2.0'' and ``Finance 4.0'' (see~\cite{capitalism20,beyond:superintelligence,qualified:money,blog:why:need} for details).

A special case of a multidimensional incentive system is ``qualified money''. The concept was first introduced by Dirk Helbing in~\cite{thinking:ahead,qualified:money}. 
Instead of a scalar (one-dimensional) quantity, like the Euro or any other currency, money could be multidimensional and earn its own reputation. 
To illustrate this, consider the example that there were two dimensions of money. 
By law, the first could only be invested into real values, but not into financial products. Instead, the latter dimension could. There would be an exchange rate (and cost) to convert one dimension into the other. As a consequence, the ECB could increase the amount of money for real investments directly, hence avoiding the problem mentioned earlier. In other words, the decision space on which institutions like the ECB can act
would considerably increase without them acting outside of their mandate. 
Qualified money, which could be realized in a Bitcoin-like~\cite{bitcoin} way\footnote{That means, transactions are transparent. It is important to have a dimension of qualified money which cannot be tracked, and this dimension should loose value more rapidly to incentive spending it soon. See~\cite{qualified:money} for details.}, could earn its own reputation depending on how and where it was created and what businesses it supports. The reputation then can give the money more or less value, which can lead to a more sustainable economy as sustainability would become measurable and transparent to individuals (for details see~\cite{capitalism20,qualified:money}).
The concept of qualified money is not limited to the above described two dimensions. Instead, everything people care about can be represented by a dimension in the currency vector.
As we explain in the following, one dimension of qualified money could be socio-digital capital that can be acquired in the digital world. 

Modern information and communication technologies play an important role 
in facing todays challenges. 
Indeed, nowadays the digital and physical world are strongly interdependent and cannot be treated in isolation any more~\cite{helbing2013globally}. 
The huge success of Web 2.0 and online social networks is changing the way humans interact at a global scale. They promote cooperation and collaboration on unprecedented scales, but at the same time powerful monopolies of information have the power to alter individuals' emotions and decisions~\cite{Epstein18082015,Bond2012}. Supercomputers nowadays perform a large fraction of all financial transactions, hence influencing the prices of important commodities, which can lead to starving, conflicts, war, etc. Information and communication technologies thus are both a crucial part of the problems society has to solve as well as a fundamental and promising piece of the solution.

\section{Decentralized information architectures and qualified money: A Social Bitcoin}

\subsection{Decentralized architectures}

The existence of powerful monopolies of information like big IT-companies or even some governments can lead to the loss of control by individuals, companies, or states. 
Besides, the economic damage attributed to cybercrime is growing exponentially and is estimated to reach 2.1 trillion dollars in 2019~\cite{cybercrime}. 
Hence, it is time to design more resilient information and communication technologies that---by design---cannot be exploited by single entities. Decentralized architectures naturally provide these benefits~\cite{helbing:digital_democracy,Contreras11122015,isocial}.

\subsection{Social Bitcoins and Web 4.0}
\label{sec_socbit}

As explained earlier, the main idea behind qualified money is to price a broader spectrum of externalities.
This means that it can be applied to, for instance, information. This can be realized in many different ways. The exact details would probably emerge in a self-organized way, depending on choices and preferences of individuals. But how could such a system look like and what benefits would it provide? Here, we discuss a possible vision in which one dimension of qualified money, socio-digital capital, can be priced in terms of Social Bitcoins that can be mined using online social networks and digital infrastructures. It is impossible to foresee the exact details of such a system, nevertheless, in the following we will sketch a possible vision of a future Internet and digital world in which individuals perform the routing of messages and information using their social contacts and technological connections rather than relying on service providers.  

The use of the Internet has changed fundamentally since its invention. At first, it was a collection of static web pages. Then,
Web 2.0 emerged as ``a collaborative medium, a place where we [could] all meet and read and write''~\cite{wiki:web20}. Consequently, Web 3.0 constitutes a ``Semantic Web''~\cite{wiki:web20}, where data can be processed by machines. Let us refer to a digital world in which information is managed in a bottom-up way, free of central monopolies in control of the vast majority of information, as Web 4.0\footnote{In~\cite{web40} Web 4.0 is described as follows: ``Web 4.0 will be as a read-write-execution-concurrency web with intelligent interactions, but there
is still no exact definition of it. Web 4.0 is also known as symbiotic web in which human mind
and machines can interact in symbiosis.''}. A digital democracy~\cite{helbing:digital_democracy}, if you will. 
Assume that this digital world was composed of many interacting, decentralized systems, which---in the absence of central control---compete for the attention of individuals~\cite{ecology20,worldmodel}. As we explain in Sec.~\ref{sec_model}, such a state is possible but fragile. Now assume that, in the future of the Internet, each individual routes information using their social and technological connections rather than relying on service providers\footnote{It is important to note that this new type of information routing requires efficient and secure encryption to ensure privacy of individuals, whenever they wish so.}. In decentralized architectures, this task has to be performed relying only on local knowledge. As shown in~\cite{geometry:multilayer}, this type of routing can be performed very efficiently and---most importantly---can be perfected if individuals actively use multiple networks simultaneously\footnote{This is only the case if the different networks are related such that they exhibit \textit{geometric correlations}. As shown in~\cite{geometry:multilayer}, real systems obey this condition. }. 
This fact constitutes an important starting point to design appropriate incentives to sustain digital democracy.

Assume that individuals could earn Social Bitcoins by routing information in the way explained above. These Social Bitcoins would form a dimension of qualified money~\cite{thinking:ahead,capitalism20,qualified:money} and could (with some additional cost) be exchanged and hence converted into other dimensions of the currency vector.
Their exchange rate would depend on the trust individuals have in the system and how much they value their socio-digital environment.

\begin{figure}[t]
\centering
 \includegraphics[width=1\linewidth]{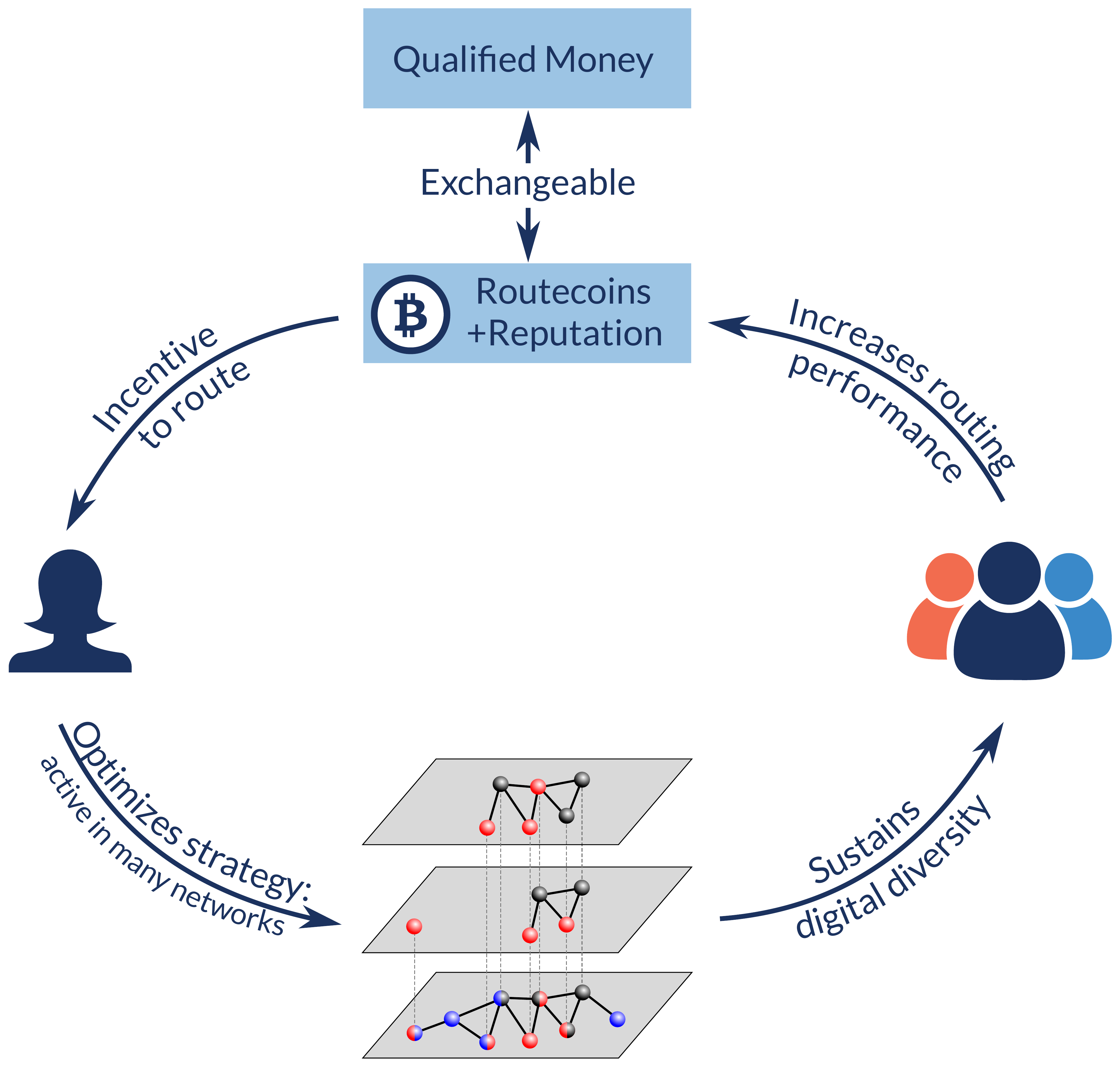}
 \caption{\textbf{Illustration of the incentive to mine Social Bitcoins.} Social Bitcoins form one dimension of qualified money and can be mined by performing  search and navigation tasks using social and technological connections
 in a future digital world. Hence, acquiring Social Bitcoins constitutes an incentive to perform routing. Individuals optimize their strategy to route information and will be active simultaneously in more networks, as this (among other aspects~\cite{nav:game}) increases routing performance~\cite{geometry:multilayer}. 
Performing routing in less active networks could increase the reputation of the mined Social Bitcoins, providing an additional incentive to engage in less active networks.
 This then could sustain digital diversity~\cite{ecology20,worldmodel} and at the same time increase the performance of routing.   
 \textit{Icon credits: Mourad Mokrane, lastspark, and Joshua Jones from thenounproject.com (CC BY 3.0).}
 \label{fig_scheme}}
\end{figure}

The important point is that now individuals have an incentive to route information (``mining'' Social Bitcoins). As a consequence, individuals will optimize to some extend their capabilities to perform this action. As explained above and shown in~\cite{geometry:multilayer}, the routing success can be increased and even perfected if individuals actively use many networks simultaneously\footnote{In addition, there are other aspects individuals might optimize, see for instance~\cite{nav:game}.}. Hence, the introduction of a Social Bitcoin would constitute an incentive to be active in several networks, as illustrated in Fig.~\ref{fig_scheme}. 
In addition, search and navigation tasks taking place in less active networks could increase the reputation of the mined Social Bitcoins, providing further incentive to engage in less active networks.
In other words, sustainability in the digital world could be priced and would become transparent to individuals who then could adjust their behavior accordingly. 
Importantly, as we explain in detail in the following, this optimization could make digital diversity robust and sustainable.

\subsection{How a Social Bitcoin could sustain digital diversity}
\label{sec_model}

Here, we present a mathematical model to illustrate the potential effect of a Social Bitcoin.
As mentioned earlier, many digital services compete for the attention of individuals. In this context, the attention of users
can be considered a scarce resource and hence the digital world forms a complex ecosystem in which networks represent competing species.
A concise description of the digital ecology was developed in~\cite{ecology20}.
In a nutshell, multiple online social networks compete for the attention of individuals in addition to obeying their intrinsic evolutionary dynamics. This dynamics is given by two main mechanisms, the influence of mass media and a viral spreading dynamics acting on top of pre-existing underlying offline social networks~\cite{our:model}.
Importantly, the parameter that quantifies the strength of viral spreading, $\lambda$, determines the final fate of the network. If $\lambda$ is below a critical value $\lambda_c$, the network will eventually become entirely passive, with corresponds to the death of the network. On the other hand, for $\lambda > \lambda_c$, the activity of the network is sustained~\cite{our:model,Bruno:Ribeiro}. 
The competition between multiple networks can be modeled assuming that more active networks are more attractive to users. 
Hence, the total virality, which reflects the overall involvement of individuals in online social networks, is distributed between the different networks as a function of their activities. More active networks obtain a higher share of the virality, which then makes these networks more active. Note that this induces a rich-get-richer effect. Interestingly, despite this positive feedback loop, diminishing returns induced by the network dynamics allows for a stable coexistence (digital diversity) of several networks in a certain parameter range (we refer the reader to~\cite{ecology20} for details).

The system can be described by the following meanfield equations\footnote{In the framework of~\cite{ecology20}, 
these equations are the result of taking the limit of $\nu \rightarrow \infty$, where $\nu$ describes the ratio between the rate at which the viral spreading and the influence of mass media occur. As shown in~\cite{ecology20}, taking this limit has no impact on the stability of the system.}
\begin{equation}
 \dot{\rho}^\mathrm{a}_{i}  = \rho^\mathrm{a}_{i} \biggr\{ \lambda \km \omega_i(\gvec{\rho}^\mathrm{a})  \left[ 1 - 
\rho^\mathrm{a}_{i} \right] -1 \biggl\} \eqcomma \quad i = 1, \dots, n \eqcomma
\label{dynamicalsystem}
\end{equation}
where $\rho^\mathrm{a}_i$ denotes the fraction of active users in network $i$, $\lambda$ is the total virality mentioned earlier, and $\km$ denotes the mean degree of the network, i.e. the average number of connections each node has. 
The weights $\omega_i(\gvec{\rho}^a)$ depend on the activities in all networks, $\gvec{\rho}^\mathrm{a} = \left(\rho_1^\mathrm{a},\rho_2^\mathrm{a}, \dots, \rho_n^\mathrm{a}  \right) $, and govern the distribution of virality between different networks. In~\cite{ecology20} we used
$
 \omega_i(\gvec{\rho}^\mathrm{a}) = \left[ \rho_i^\mathrm{a} \right]^\sigma /
 \sum_{j=1}^{n} \left[ \rho_j^\mathrm{a} \right]^\sigma
$,
where $\sigma$ denotes the activity affinity that quantifies how much more prone individuals are to engage in more active networks. 

As mentioned earlier, assume that the introduction of Social Bitcoins incentivizes users to simultaneously use multiple networks in order to increase their capabilities to successfully perform search and navigation tasks and hence increase their expected payoff. The exact form of this incentive depends on the details of the implementation of the systems' architectures and Social Bitcoins, which comprises an interesting future research direction.
Here, we model the additional tendency of individuals to engage in multiple (and less active) networks by shifting the weight of the distribution of the virality towards networks with lower activity, hence hindering the rich-get-richer effect described earlier.  
In particular, let us consider the following form of the weight function,
\begin{equation}
 \omega_i(\gvec{\rho}^\mathrm{a}) = \underbrace{\frac{\left[ \rho_i^\mathrm{a} \right]^\sigma}{\sum_{j=1}^{n_l} \left[ \rho_j^\mathrm{a} \right]^\sigma}}_{\text{rich-get-richer~\cite{ecology20}}}
 + \underbrace{\xi (\left<\gvec{\rho}^a\right>-\rho_i^\mathrm{a})}_{\text{Social Bitcoin incentive}}
 \eqcomma
 \label{eq_weights}
\end{equation}
where $\xi$ is a parameter proportional to the value of Social Bitcoins and $\left<\gvec{\rho}^a\right> = \frac{1}{n} \sum_{i=1}^n \rho_i^\mathrm{a}$ denotes the mean activity among all networks.

\begin{figure}[t]
\centering
\includegraphics[width=1\linewidth]{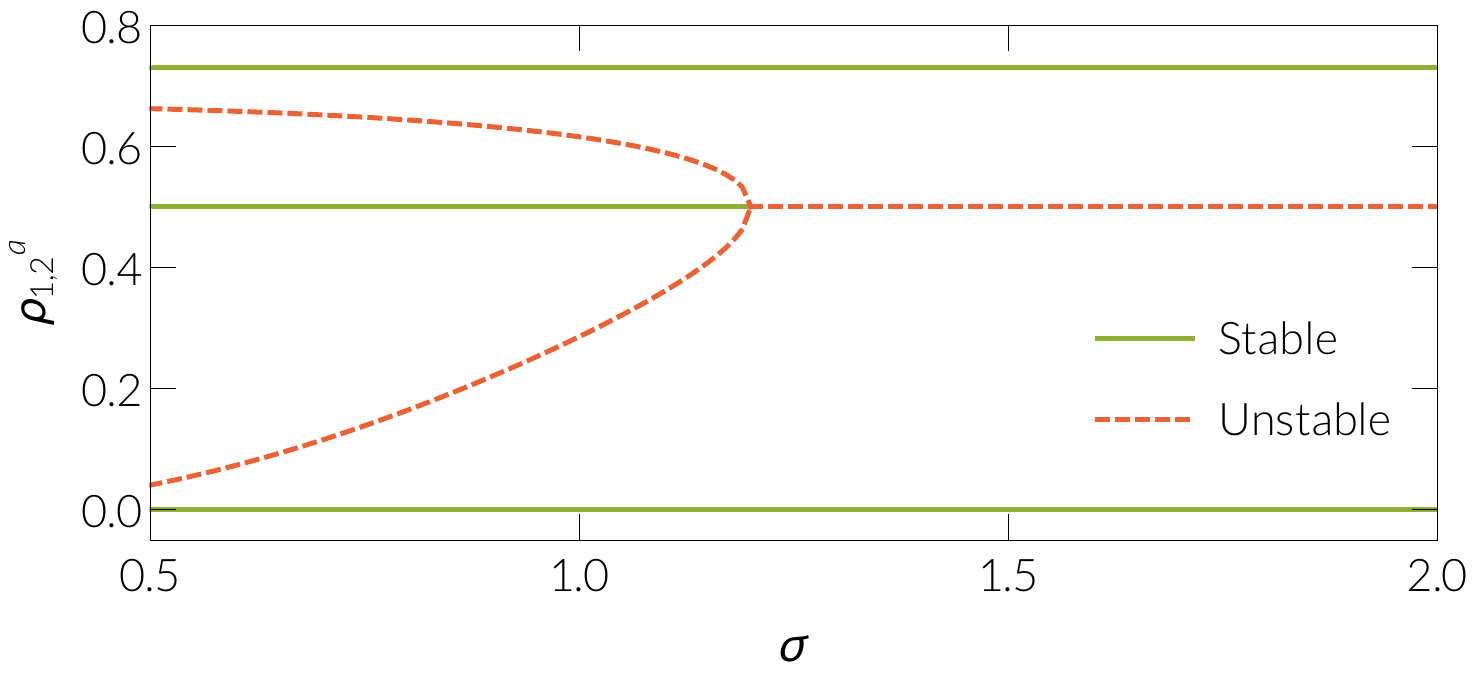}
\includegraphics[width=1\linewidth]{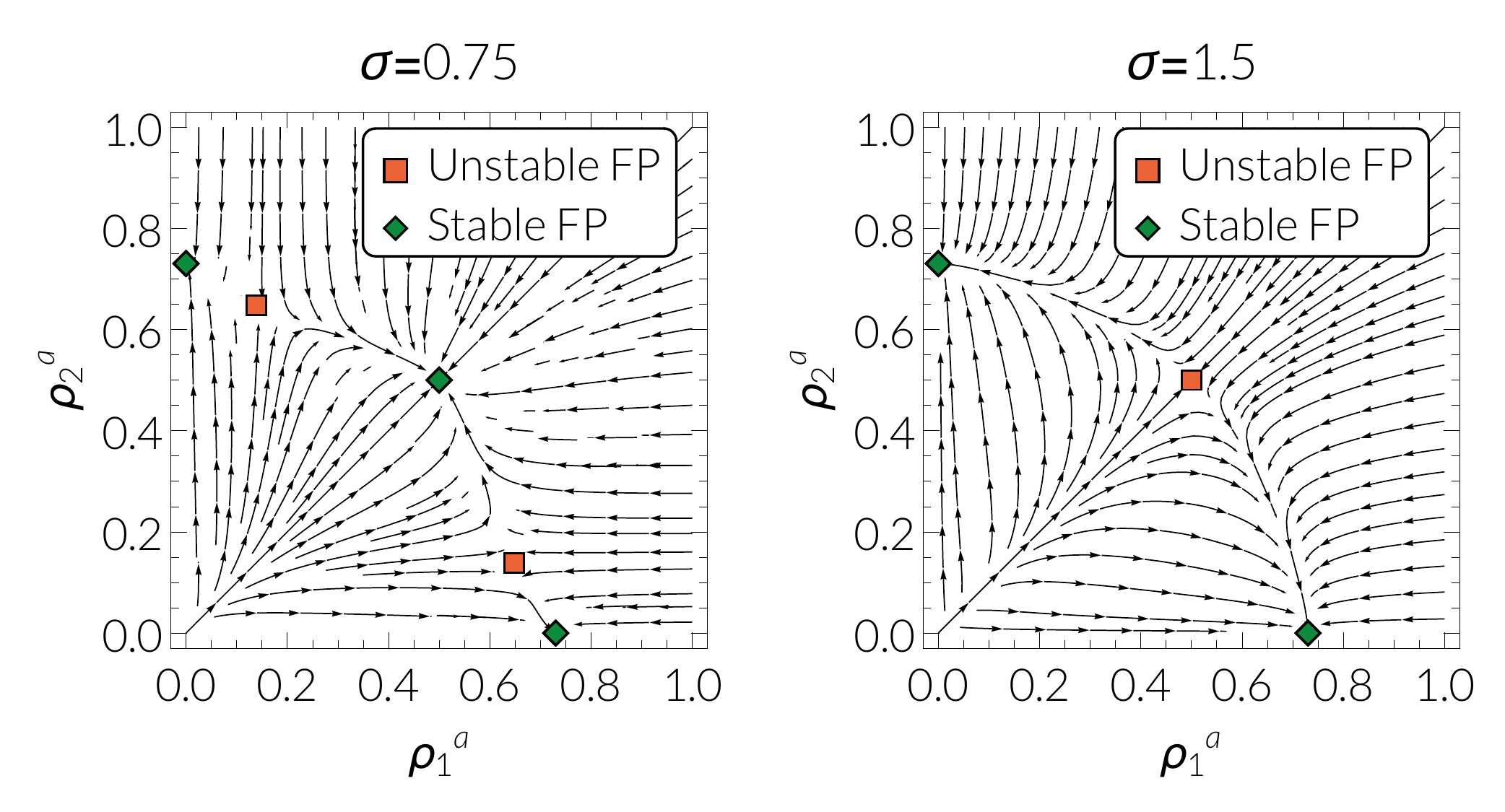}
\includegraphics[width=1\linewidth]{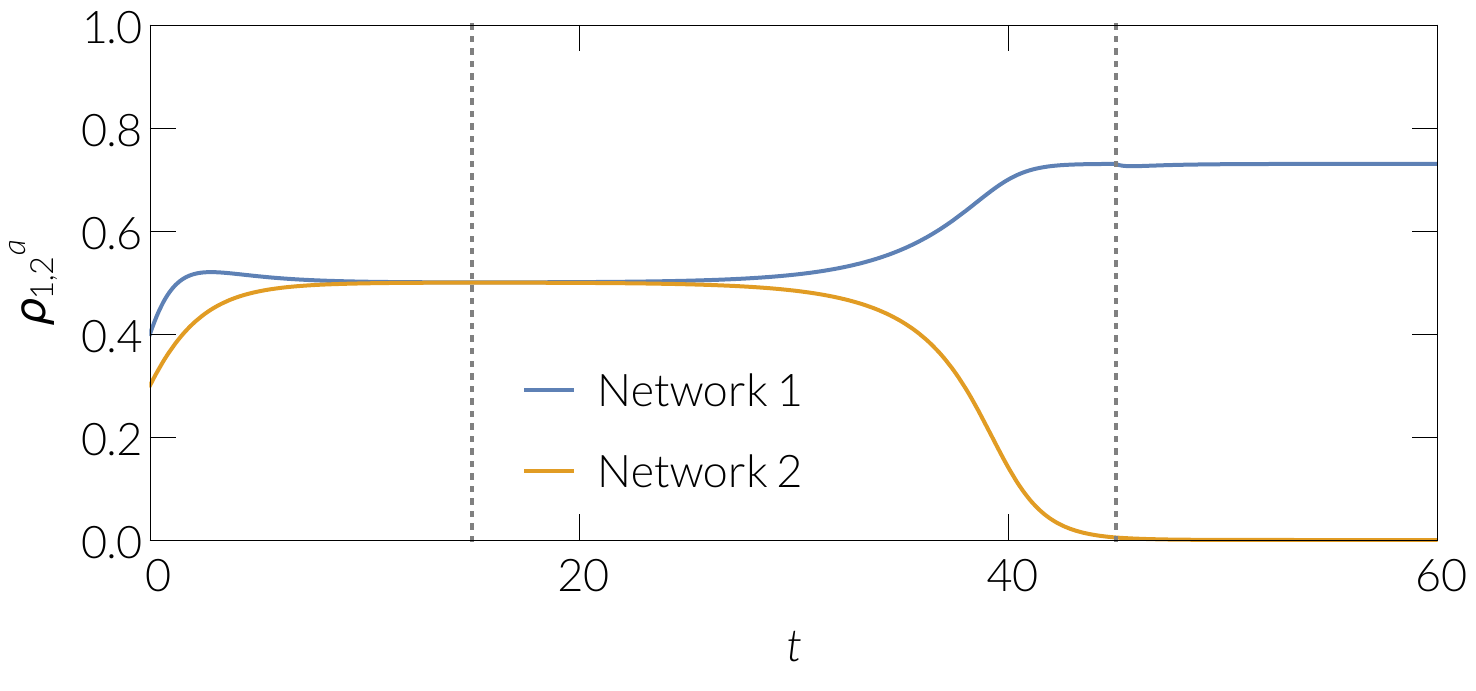}
\caption{
\textbf{Fragility of digital diversity.} Here, we consider two networks,
$\lambda \km = 2$ and $\xi = 0.2$.
\textbf{Top:} Bifurcation diagram (subcritical pitchfork bifurcation). $\rho_i$ denotes the fraction of active users in network $i$. Green solid lines represent stable solutions and red dashed lines correspond to unstable fixed points. 
\textbf{Middle:} Stream line plots for $\sigma=0.75$ (left) and $\sigma=1.5$ (right). 
\textbf{Bottom:} Evolution of the system for initial conditions $\rho_1 = 0.4$, $\rho_2 = 0.3$. 
For $15 \leq t < 45$ (between the dashed lines) we set $\sigma = 1.5$ and otherwise we set $\sigma = 0.75$.
\label{fig_bif}}
\end{figure}

The effect of the inclusion of the new term (``Social Bitcoin incentive'') in Eq.~\eqref{eq_weights} can change the behavior of the system dramatically if $\xi$ is large enough, which we illustrate\footnote{Here we present only a brief discussion of the dynamical system given by Eqns.~\eqref{dynamicalsystem} and~\eqref{eq_weights}. A more detailed analysis and the investigation of different forms of the incentive term in Eq.~\eqref{eq_weights} is left for future research.} for two competing networks. Let us first consider the case of $\xi = 0.2$. In this case, the qualitative behavior of the system is similar to the one in absence of Social Bitcoins as described in~\cite{ecology20}. 
Below a critical value of the activity affinity, $\sigma < \sigma_c$, coexistence is possible (solid green central branch in Fig.~\ref{fig_bif}~(top) \& central green diamond in Fig.~\ref{fig_bif}~(middle, left)), but---once lost---cannot be recovered. 

To illustrate this, assume that we start with $\sigma < \sigma_c$ and the system approaches the coexistence solution (central green diamond in Fig.~\ref{fig_bif}~(middle, left)). Then, we change $\sigma$ to some value larger than $\sigma_c$. Hence, the coexistence solution becomes unstable and the system eventually approaches the solution where either $\rho_1 = 0$ or $\rho_2=0$ (green diamonds in Fig.~\ref{fig_bif}~(middle, right)). Now, after changing $\sigma$ back to a value below $\sigma_c$, the system does not return to the again stable coexistence state, but instead remains in the domination state, which is also stable (outer green diamonds in Fig.~\ref{fig_bif}~(middle, left)). This example is illustrated in Fig.~\ref{fig_bif}~(bottom), where we explicitly show the evolution of the fraction of active users for both networks\footnote{Note that here we describe an idealized system without noise. Noise in real systems would speed up significantly the separation of the trajectories in Fig.~\ref{fig_bif}~(bottom) shortly after the first dashed gray line.}. 
To conclude, the system is fragile in the sense that an irreversible loss of digital diversity is possible---similar to the loss of biodiversity.

\begin{figure}[t]
\centering
\includegraphics[width=1\linewidth]{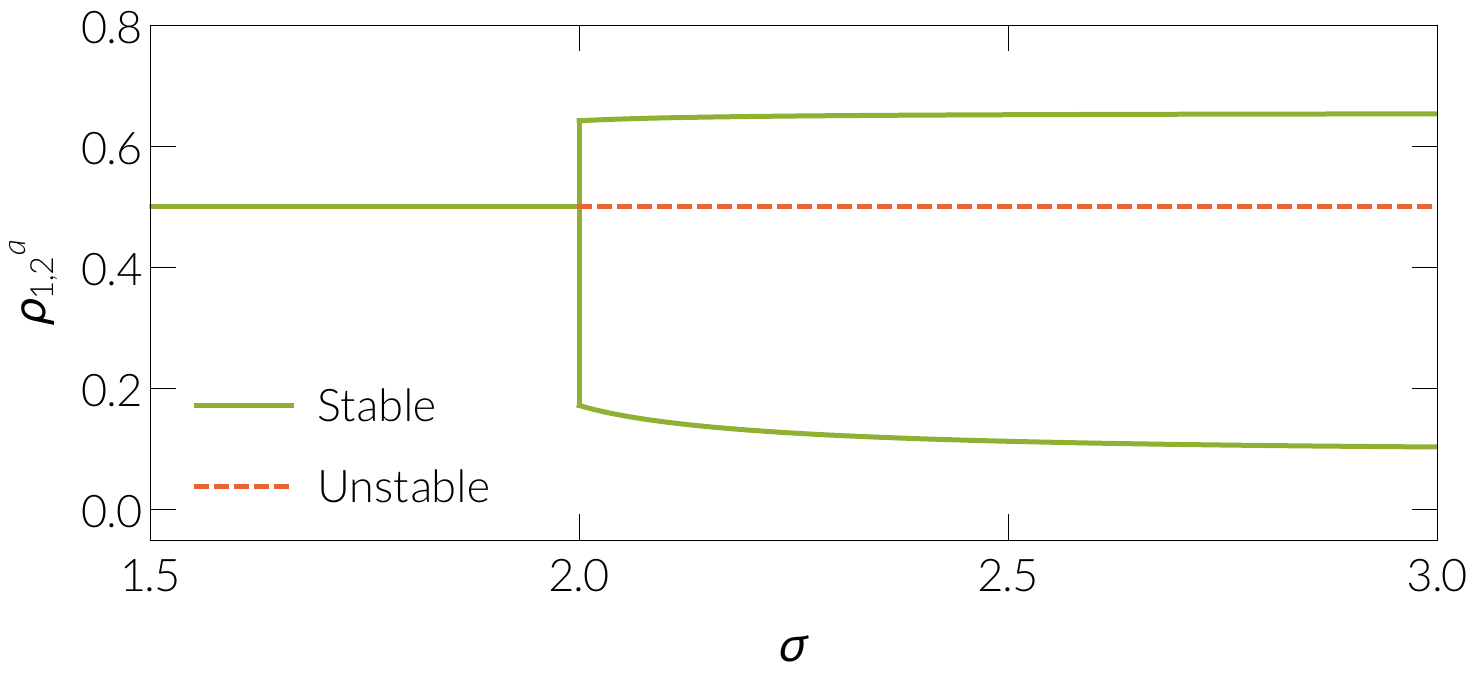}
\includegraphics[width=1\linewidth]{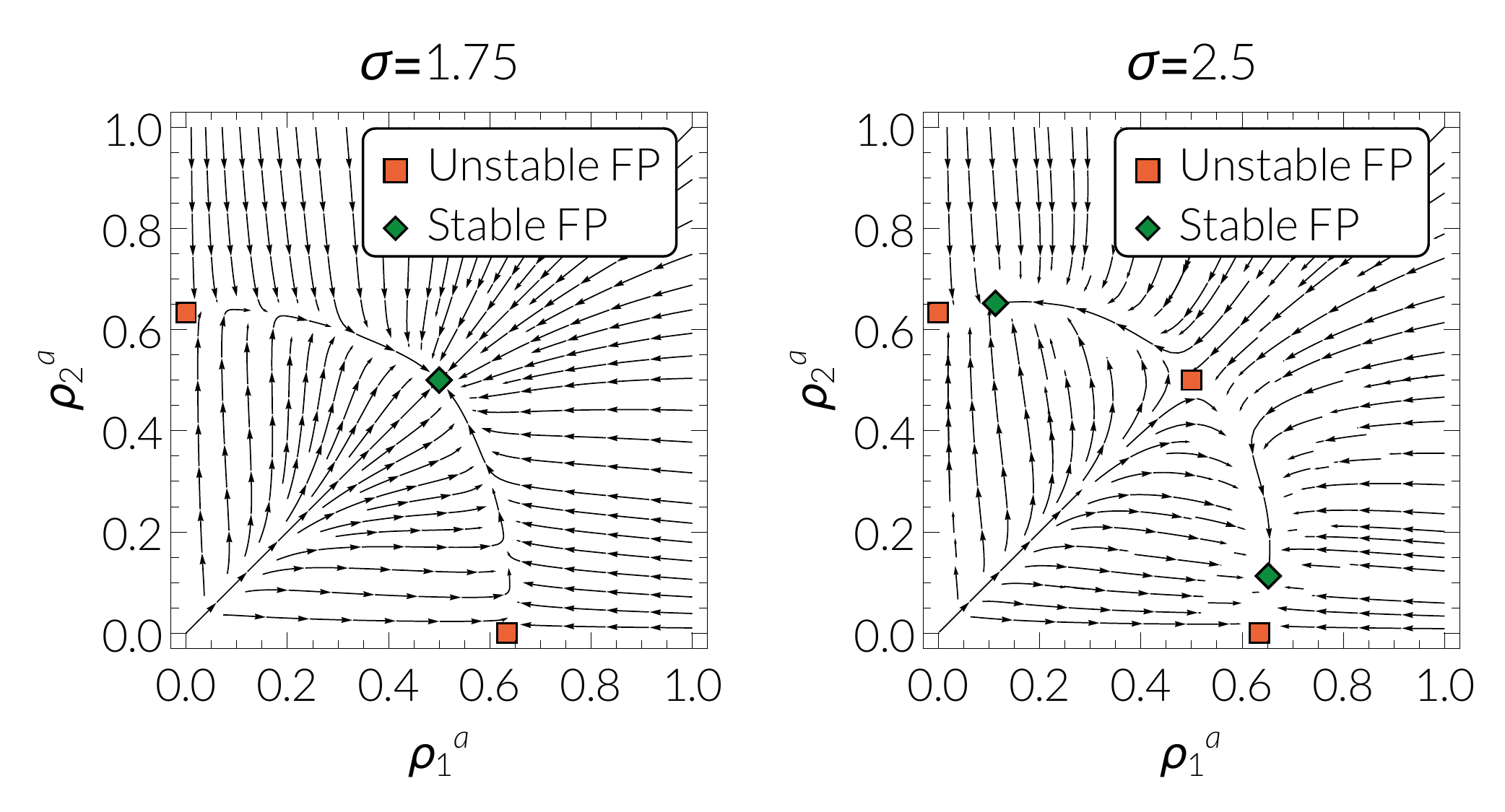}
\includegraphics[width=1\linewidth]{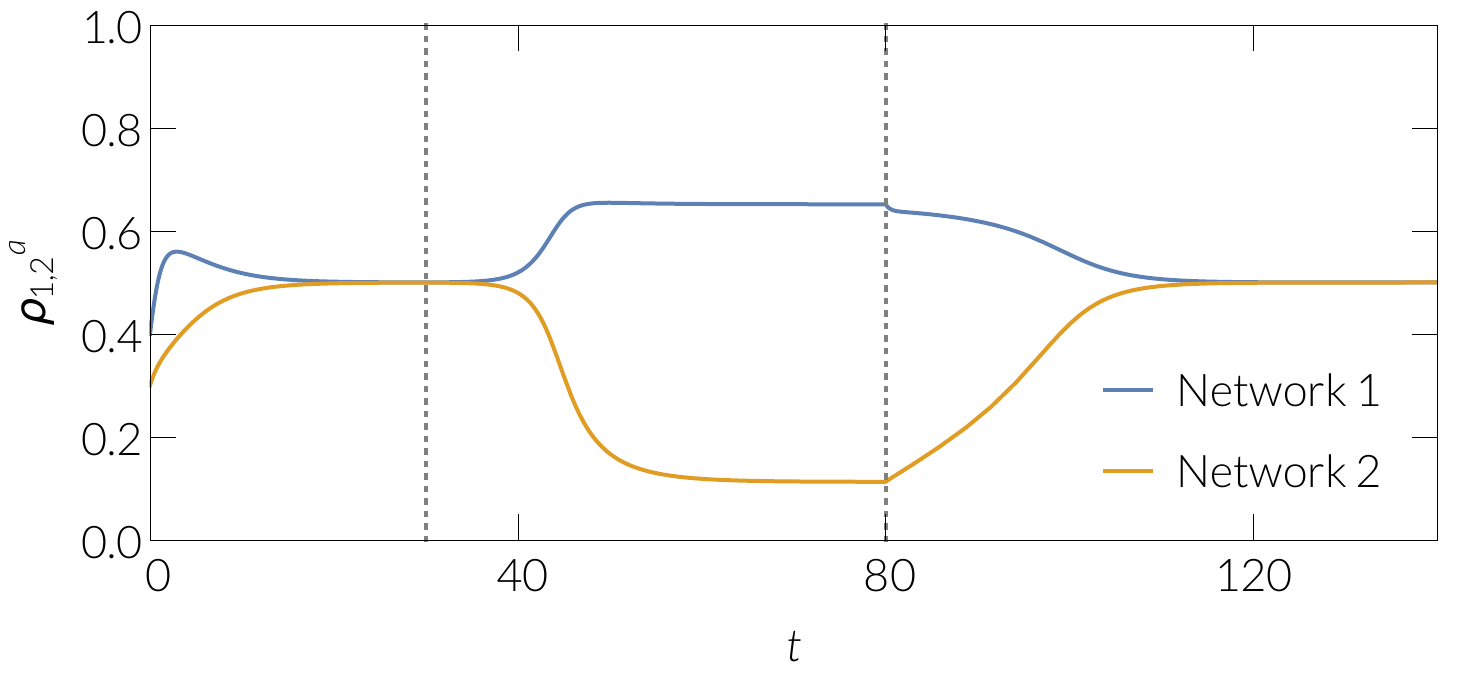}
\caption{\textbf{
Robustness of digital diversity.} Here, we consider two networks,
$\lambda \km = 2$ and $\xi = 1.0$.
\textbf{Top:} Bifurcation diagram. $\rho_i$ denotes the fraction of active users in network $i$. Green solid lines represent stable solutions and red dashed lines correspond to unstable fixed points. For better readability, here we do not show the unstable fixed points for $\rho_1=0$ and $\rho_2=0$. 
\textbf{Middle:} Stream line plots for $\sigma=0.75$ (left) and $\sigma=2.5$ (right). 
\textbf{Bottom:} Evolution of the system for initial conditions $\rho_1 = 0.4$, $\rho_2 = 0.3$.
For $30 \leq t < 80$ (between the dashed lines) we set $\sigma = 2.5$ and otherwise we set $\sigma = 1.75$.
\label{fig_bif2}}
\end{figure}

Interestingly, for a higher value of $\xi$ the behavior of the system differs dramatically, which we illustrate here for $\xi=1$. The solution corresponding to equal coexistence of two networks, hence $\rho_1 = \rho_2 \neq 0$, is stable as before for values of $\sigma$ below some critical value $\sigma_c$. However, in this regime now the domination solutions ($\rho_1 = 0$ or $\rho_2 = 0$, denoted by the red squares in Fig.~\ref{fig_bif2}~(middle, left)) are unstable. This means that, independently from the initial conditions, in this regime the system always approaches the coexistence solution. For $\sigma > \sigma_c$ the equal coexistence solution becomes unstable and new stable solutions emerge (green diamonds in Fig.~\ref{fig_bif2}~(middle, right)). These unequal coexistence solutions correspond to the case that one network has a significantly higher activity than the other, but the activities of both networks are sustained. 
Let us again consider the explicit example of two networks and start with $\sigma < \sigma_c$. The system approaches the equal coexistence solution (green square in Fig.~\ref{fig_bif2}~(middle, left)). Then, we change $\sigma$ to some value above $\sigma_c$. Now, the system approaches the unequal coexistence solution (green diamonds in Fig.~\ref{fig_bif2}~(middle, right)), but now the activity in both networks is sustained. By lowering $\sigma$ again below $\sigma_c$, the system recovers the equal coexistence solution, in contrast to the previous case. 
This example is shown in Fig.~\ref{fig_bif2}~(bottom) where we present the fraction of active users in both networks.
To conclude, in contrast to the case discussed before, the system is robust in the sense that an irreversible loss of digital diversity cannot occur. 

To sum up, the introduction of a multidimensional incentive system in which one dimension represents socio-digital capital in terms of Social Bitcoins that can be mined by performing search and navigation tasks in a future digital world can make digital diversity robust---given that the value of Social Bitcoins is high enough.

\section{Outlook and future research directions}

A multidimensional financial system offers manifold success opportunities for individuals, companies, and states. Top-down control alone is destined to fail in a hyperconnected world. Hence, we need a new approach that incorporates the bottom-up empowerment of society and the right incentives and feedback mechanisms to promote creativity and innovations.
The initiative ``A nation of makers''~\cite{nation:of:makers} in the US as well as the rise of citizen science~\cite{citizen:science} constitute promising starting points for such a development. Nevertheless, increasing financial instabilities emphasize the pressing need to redesign certain aspects of the financial system, hence the urge to create ``Finance 4.0''~\cite{qualified:money}.

In this perspective, multiple monetary dimensions could represent different externalities (negative ones like noise, environmental damage, etc. and positive ones such as recycling of resources, cooperation, creation of new jobs and so on). Building on this framework, appropriate feedback and coordination mechanisms could increase resource efficiency and lead to a more sustainable, circular, cooperative economy. 
This can be achieved in a bottom-up way in terms of an improved version of capitalism based on the abilities of self-organization intrinsically present in dynamical complex systems by accounting for externalities in a multidimensional incentive system. The Internet of Things and the blockchain technology underlying the Bitcoin architecture provide the technological requirements to realize ``Finance 4.0'' and ``Capitalism 2.0'' based on knowledge from the science of complex systems~\cite{capitalism20,beyond:superintelligence,qualified:money,iot:hand,blog:why:need}. 

Nowadays, the digital and physical world are strongly interdependent. We have presented an example how a multidimensional incentive system, in particular a Social Bitcoin generated in a bottom-up way by performing search and navigation tasks in a possible future digital world can sustain digital diversity, which is essential for the freedom of information. Furthermore, a diverse digital landscape is expected to create business opportunities for individuals and companies~\cite{new:economy,beyond:superintelligence,iot:hand} facing the disappearance of half of todays jobs~\cite{Frey13thefuture}. 
The price of Social Bitcoins is crucial for the desired effect of sustaining digital diversity. This price, however, is determined dynamically by the market 
and may depend on other dimensions of the currency system.
The development of a concise and general theory of this system and possible implementations comprise interesting future research directions.

\begin{acknowledgments}
K-K. Kleineberg acknowledges support by the European Commission through the Marie Curie ITN ``iSocial'' grant no.\ PITN-GA-2012-316808.
\end{acknowledgments}


\begin{thebibliography}{10}
\expandafter\ifx\csname url\endcsname\relax
  \def\url#1{\texttt{#1}}\fi
\expandafter\ifx\csname urlprefix\endcsname\relax\def\urlprefix{URL }\fi
\providecommand{\bibinfo}[2]{#2}
\providecommand{\eprint}[2][]{\url{#2}}

\bibitem{helbing2013globally}
\bibinfo{author}{Helbing, D.}
\newblock \bibinfo{title}{Globally networked risks and how to respond}.
\newblock \emph{\bibinfo{journal}{Nature}} \textbf{\bibinfo{volume}{497}},
  \bibinfo{pages}{51--59} (\bibinfo{year}{2013}).

\bibitem{BPPSH10}
\bibinfo{author}{Buldyrev, S.~V.}, \bibinfo{author}{Parshani, R.},
  \bibinfo{author}{Paul, G.}, \bibinfo{author}{Stanley, H.~E.} \&
  \bibinfo{author}{Havlin, S.}
\newblock \bibinfo{title}{Catastrophic cascade of failures in interdependent
  networks}.
\newblock \emph{\bibinfo{journal}{Nature}} \textbf{\bibinfo{volume}{464}},
  \bibinfo{pages}{1025--1028} (\bibinfo{year}{2010}).

\bibitem{interaction:support}
\bibinfo{author}{Helbing, D.}
\newblock \bibinfo{title}{Interaction support processor}
  (\bibinfo{year}{2015}).
\newblock
  \bibinfo{note}{\url{https://patentscope.wipo.int/search/en/detail.jsf?docId=WO2015118455}}.

\bibitem{new:economy}
\bibinfo{author}{Helbing, D.}
\newblock \bibinfo{title}{Why we need a new economy to survive}
  (\bibinfo{year}{2016}).
\newblock \bibinfo{note}{\url{https://www.youtube.com/watch?v=OV_b3b_Spow}}.

\bibitem{helbing:self}
\bibinfo{author}{Helbing, D.}
\newblock \emph{\bibinfo{title}{Social Self-Organization}}
  (\bibinfo{publisher}{Springer}, \bibinfo{year}{2012}).

\bibitem{blog:self-organized:society}
\bibinfo{author}{Helbing, D.}
\newblock \bibinfo{title}{The self-organizing society - taking the future in
  our hands} (\bibinfo{year}{2015}).
\newblock
  \bibinfo{note}{\url{http://futurict.blogspot.com.es/2015/01/the-self-organizing-society-taking.html}}.

\bibitem{page:difference}
\bibinfo{author}{Page, S.~E.}
\newblock \emph{\bibinfo{title}{The Difference: How the Power of Diversity
  Creates Better Groups, Firms, Schools, and Societies}}
  (\bibinfo{publisher}{Princeton Univ. Press}, \bibinfo{year}{2008}).

\bibitem{2016man}
\bibinfo{author}{Arpe, J.}
\newblock \emph{\bibinfo{title}{To the Man with a Hammer: Augmenting the
  Policymaker's Toolbox for a Complex World}} (\bibinfo{publisher}{Bertelsmann
  Stiftung}, \bibinfo{year}{2016}).

\bibitem{change:complex}
\bibinfo{author}{Helbing, D.}
\newblock \bibinfo{title}{Implementing change in a complex world - responding
  to complexity in socio-economic systems: How to build a smart and resilient
  society?} (\bibinfo{year}{2015}).
\newblock
  \bibinfo{note}{\url{http://futurict.blogspot.com.es/2015/03/implementing-change-in-complex-world.html}}.

\bibitem{ecology20}
\bibinfo{author}{Kleineberg, K.-K.} \& \bibinfo{author}{Bogu\~n\'a, M.}
\newblock \bibinfo{title}{Digital ecology: Coexistence and domination among
  interacting networks}.
\newblock \emph{\bibinfo{journal}{Sci. Rep.}} \textbf{\bibinfo{volume}{5}},
  \bibinfo{pages}{10268} (\bibinfo{year}{2015}).

\bibitem{worldmodel}
\bibinfo{author}{Kleineberg, K.-K.} \& \bibinfo{author}{Bogu\~n\'a, M.}
\newblock \bibinfo{title}{Competition between global and local online social
  networks}.
\newblock \emph{\bibinfo{journal}{Sci. Rep.}} \textbf{\bibinfo{volume}{6}},
  \bibinfo{pages}{25116} (\bibinfo{year}{2016}).

\bibitem{PhysRevLett.59.381}
\bibinfo{author}{Bak, P.}, \bibinfo{author}{Tang, C.} \&
  \bibinfo{author}{Wiesenfeld, K.}
\newblock \bibinfo{title}{Self-organized criticality: An explanation of the 1/f
  noise}.
\newblock \emph{\bibinfo{journal}{Phys. Rev. Lett.}}
  \textbf{\bibinfo{volume}{59}}, \bibinfo{pages}{381--384}
  (\bibinfo{year}{1987}).
\newblock \urlprefix\url{http://link.aps.org/doi/10.1103/PhysRevLett.59.381}.

\bibitem{reuters:negative}
\bibinfo{author}{Reuters}.
\newblock \bibinfo{title}{Negative rates for 2-3 years become worry for banks}
  (\bibinfo{year}{Apr 7 2016}).
\newblock
  \bibinfo{note}{\url{http://www.reuters.com/article/eurozone-ecb-policy-idUSF9N12D02K}}.

\bibitem{quantitative:easing}
\bibinfo{author}{Reuters}.
\newblock
  \bibinfo{title}{http://www.reuters.com/article/us-eurozone-bonds-ecb-iduskcn0wp1o3}
  (\bibinfo{year}{Mar 23 2016}).
\newblock
  \bibinfo{note}{\url{http://www.reuters.com/article/us-eurozone-bonds-ecb-idUSKCN0WP1O3}}.

\bibitem{reuters:credibility}
\bibinfo{author}{Reuters}.
\newblock \bibinfo{title}{Ecb's credibility at stake if inflation target
  missed} (\bibinfo{year}{Apr 7 2016}).
\newblock
  \bibinfo{note}{\url{http://www.reuters.com/article/eurozone-ecb-inflation-idUSF9N13T000}}.

\bibitem{reuters:helicopter}
\bibinfo{author}{Reuters}.
\newblock \bibinfo{title}{Ecb could give 1,300 euros to bloc's citizens, nordea
  says} (\bibinfo{year}{Mar 31 2016}).
\newblock
  \bibinfo{note}{\url{http://www.reuters.com/article/ecb-policy-cash-idUSL5N1733LR}}.

\bibitem{helicopter}
\bibinfo{author}{Lynn, M.}
\newblock \bibinfo{title}{Draghi may have to throw money out of a helicopter}
  (\bibinfo{year}{March 2016}).
\newblock
  \bibinfo{note}{\url{http://www.telegraph.co.uk/business/2016/03/07/draghi-may-have-to-throw-money-out-of-a-helicopter/}}.

\bibitem{reuter:fed}
\bibinfo{author}{Reuters}.
\newblock \bibinfo{title}{Fed signals caution on rate hikes, worried by global
  growth: minutes}.
\newblock
  \bibinfo{note}{\url{http://www.reuters.com/article/us-fed-minutes-idUSKCN0X32AB}}.

\bibitem{thinking:ahead}
\bibinfo{author}{Helbing, D.}
\newblock \emph{\bibinfo{title}{Thinking Ahead - Essays on Big Data, Digital
  Revolution, and Participatory Market Society}}
  (\bibinfo{publisher}{Springer}, \bibinfo{year}{2015}).

\bibitem{capitalism20}
\bibinfo{author}{Helbing, D.}
\newblock \bibinfo{title}{From communism 2.0 to capitalism 2.0}
  (\bibinfo{year}{March 2016}).
\newblock
  \bibinfo{note}{\url{http://futurict.blogspot.com.es/2016/03/from-communism-20-to-capitalism-20.html}}.

\bibitem{beyond:superintelligence}
\bibinfo{author}{Helbing, D.}
\newblock \bibinfo{title}{Beyond superintelligence: Mastering future challenges
  with capitalism 2.0 and democracy 2.0} (\bibinfo{year}{2016}).
\newblock \bibinfo{note}{\url{https://www.youtube.com/watch?v=OV_b3b_Spow}}.

\bibitem{qualified:money}
\bibinfo{author}{Helbing, D.}
\newblock \bibinfo{title}{Qualified money: A better financial system for the
  future} (\bibinfo{year}{October 2014}).
\newblock
  \bibinfo{note}{\url{http://futurict.blogspot.com.es/2014/10/qualified-money-better-financial-system.html}}.

\bibitem{blog:why:need}
\bibinfo{author}{Helbing, D.}
\newblock \bibinfo{title}{Why we need democracy 2.0 and capitalism 2.0 to
  survive} (\bibinfo{year}{2016}).
\newblock
  \bibinfo{note}{\url{http://futurict.blogspot.com.es/2016/04/why-we-need-democracy-20-and-capitalism.html}}.

\bibitem{bitcoin}
\bibinfo{author}{Nakamoto, S.}
\newblock \bibinfo{title}{Bitcoin: A peer-to-peer electronic cash system.
  provides a portrait of what bitcoin is and how it would be implemented}.
\newblock \emph{\bibinfo{journal}{http://www.bitcoin.org/bitcoin.pdf}}
  (\bibinfo{year}{2009}).

\bibitem{Epstein18082015}
\bibinfo{author}{Epstein, R.} \& \bibinfo{author}{Robertson, R.~E.}
\newblock \bibinfo{title}{The search engine manipulation effect (seme) and its
  possible impact on the outcomes of elections}.
\newblock \emph{\bibinfo{journal}{Proceedings of the National Academy of
  Sciences}} \textbf{\bibinfo{volume}{112}}, \bibinfo{pages}{E4512--E4521}
  (\bibinfo{year}{2015}).

\bibitem{Bond2012}
\bibinfo{author}{Bond, R.~M.} \emph{et~al.}
\newblock \bibinfo{title}{{A 61-million-person experiment in social influence
  and political mobilization.}}
\newblock \emph{\bibinfo{journal}{Nature}} \textbf{\bibinfo{volume}{489}}
  (\bibinfo{year}{2012}).

\bibitem{cybercrime}
\bibinfo{title}{Cyber crime costs projected to reach \$2 trillion by 2019}
  (\bibinfo{year}{2016}).
\newblock
  \bibinfo{note}{\url{http://www.forbes.com/sites/stevemorgan/2016/01/17/cyber-crime-costs-projected-to-reach-2-trillion-by-2019/}}.

\bibitem{helbing:digital_democracy}
\bibinfo{author}{Helbing, D.} \& \bibinfo{author}{Pournaras, E.}
\newblock \bibinfo{title}{Society: Build digital democracy}.
\newblock \emph{\bibinfo{journal}{Nature}} \textbf{\bibinfo{volume}{527}},
  \bibinfo{pages}{33--34} (\bibinfo{year}{2015}).

\bibitem{Contreras11122015}
\bibinfo{author}{Contreras, J.~L.} \& \bibinfo{author}{Reichman, J.~H.}
\newblock \bibinfo{title}{Sharing by design: Data and decentralized commons}.
\newblock \emph{\bibinfo{journal}{Science}} \textbf{\bibinfo{volume}{350}},
  \bibinfo{pages}{1312--1314} (\bibinfo{year}{2015}).

\bibitem{isocial}
\bibinfo{title}{isocial: Decentralized online social networks project}
  (\bibinfo{year}{2013-2016}).
\newblock \bibinfo{note}{\url{http://isocial-itn.eu/}}.

\bibitem{wiki:web20}
 (\bibinfo{year}{2016}).
\newblock \bibinfo{note}{\url{https://en.wikipedia.org/wiki/Web_2.0}}.

\bibitem{web40}
\bibinfo{author}{Aghaei, S.}, \bibinfo{author}{Nematbakhsh, M.~A.} \&
  \bibinfo{author}{Farsani, H.~K.}
\newblock \bibinfo{title}{Evolution of the world wide web: from web 1.0 to web
  4.0}.
\newblock \emph{\bibinfo{journal}{International Journal of Web \& Semantic
  Technology (IJWesT)}} .

\bibitem{geometry:multilayer}
\bibinfo{author}{Kleineberg, K.-K.}, \bibinfo{author}{Boguñá, M.},
  \bibinfo{author}{Ángeles Serrano, M.} \& \bibinfo{author}{Papadopoulos, F.}
\newblock \bibinfo{title}{Hidden geometric correlations in real multiplex
  networks}.
\newblock \emph{\bibinfo{journal}{Nature Physics}} \bibinfo{pages}{doi:
  10.1038/NPHYS3782} (\bibinfo{year}{2016}).

\bibitem{nav:game}
\bibinfo{author}{Guly\'{a}s, A.}, \bibinfo{author}{B\'{\i}r\'{o}, J.~J.},
  \bibinfo{author}{K\H{o}r\"{o}si, A.}, \bibinfo{author}{R\'{e}tv\'{a}ri, G.}
  \& \bibinfo{author}{Krioukov, D.}
\newblock \bibinfo{title}{{Navigable networks as Nash equilibria of navigation
  games}}.
\newblock \emph{\bibinfo{journal}{Nature Communications}}
  \textbf{\bibinfo{volume}{6}}, \bibinfo{pages}{7651+} (\bibinfo{year}{2015}).

\bibitem{our:model}
\bibinfo{author}{Kleineberg, K.-K.} \& \bibinfo{author}{Bogu\~n\'a, M.}
\newblock \bibinfo{title}{Evolution of the digital society reveals balance
  between viral and mass media influence}.
\newblock \emph{\bibinfo{journal}{Phys Rev X}} \textbf{\bibinfo{volume}{4}},
  \bibinfo{pages}{031046} (\bibinfo{year}{2014}).

\bibitem{Bruno:Ribeiro}
\bibinfo{author}{Ribeiro, B.}
\newblock \bibinfo{title}{Modeling and predicting the growth and death of
  membership-based websites}.
\newblock \emph{\bibinfo{journal}{International World Wide Web Conference}}
  (\bibinfo{year}{2014}).

\bibitem{nation:of:makers}
\bibinfo{title}{A nation of makers (the white house)} (\bibinfo{year}{2016}).
\newblock \bibinfo{note}{\url{https://www.whitehouse.gov/nation-of-makers}}.

\bibitem{citizen:science}
\bibinfo{author}{Hand, E.}
\newblock \bibinfo{title}{Citizen science: People power}.
\newblock \emph{\bibinfo{journal}{Nature}} \textbf{\bibinfo{volume}{466}},
  \bibinfo{pages}{685--687} (\bibinfo{year}{2010}).

\bibitem{iot:hand}
\bibinfo{author}{Helbing, D.}
\newblock \bibinfo{title}{How the internet of things can make the invisible
  hand work and societies thrive} (\bibinfo{year}{2016}).
\newblock
  \bibinfo{note}{\url{http://futurict.blogspot.com.es/2016/02/how-internet-of-things-can-make.html}}.

\bibitem{Frey13thefuture}
\bibinfo{author}{Frey, C.~B.} \emph{et~al.}
\newblock \bibinfo{title}{The future of employment: how susceptible are jobs to
  computerisation?}  (\bibinfo{year}{2013}).
\newblock
  \bibinfo{note}{\url{http://www.oxfordmartin.ox.ac.uk/downloads/academic/The_Future_of_Employment.pdf}}.

\end{thebibliography}
\end{document}